\documentstyle[12pt,graphicx,amsmath]{article}
\newlength{\pubnumber} 

\catcode`\@=11
\@addtoreset{equation}{section}

\def\section{\@startsection{section}{1}{\z@}{3.5ex plus 1ex minus .2ex}
 {2.3ex plus .2ex}{\large\bf}}
\def\subsection{\@startsection{subsection}{2}{\z@}{2.3ex plus .2ex}
 {2.3ex plus .2ex}{\bf}}

\newcommand{\oo}[2]{\left(#1\left|#2\right.\right)}
\newcommand{\ba}{\begin{eqnarray}}
\newcommand{\ea}{\end{eqnarray}}
\begin{document}

\begin{titlepage}
\samepage{
\setcounter{page}{1}
\rightline{LPTENS--07/55}
\rightline{LTH--778}
\rightline{December 2007}

\vfill
\begin{center}
 {\Large \bf
Spinor--Vector Duality \\ in \\ $N=2$ Heterotic String Vacua
}
\vspace{1cm}
\vfill {\large Alon E. Faraggi$^{1}$,
Costas Kounnas$^{2}$\footnote{Unit\'e Mixte de Recherche
(UMR 8549) du CNRS et de l'ENS
 associ\'e¨e a l'universit\'e¨ Pierre et Marie Curie}
 and
John Rizos$^{3}$}\\
\vspace{1cm}
{\it $^{1}$ Dept.\ of Mathematical Sciences,
             University of Liverpool,
         Liverpool L69 7ZL, UK\\}
\vspace{.05in}
{\it $^{2}$ Lab.\ Physique Th\'eorique,
Ecole Normale Sup\'erieure, F--75231 Paris 05, France\\}
\vspace{.05in}
{\it $^{3}$ Department of Physics,
              University of Ioannina, GR45110 Ioannina, Greece\\}
\vspace{.025in}
\end{center}
\vfill
\begin{abstract}
Classification of the $N=1$ space--time supersymmetric
fermionic $Z_2\times Z_2$
heterotic--string vacua with symmetric internal shifts, revealed
a novel spinor--vector
duality symmetry over the entire space of vacua,
where the $S_t\leftrightarrow V$
duality interchanges the spinor plus anti--spinor representations
with vector representations.
In this paper we demonstrate that the spinor--vector
duality exists also in fermionic $Z_2$ heterotic string models,
which preserve $N=2$ space--time supersymmetry. In this case
the interchange is between spinorial and vectorial representations
of the unbroken $SO(12)$ GUT symmetry.
We provide a general algebraic proof for the
existence of the $S_t\leftrightarrow V$ duality map.
We present a novel basis to generate the free fermionic models
in which the ten dimensional gauge degrees of freedom are grouped
into four groups of four, each generating an $SO(8)$ modular block.
In the new basis the GUT symmetries are produced by generators arising
from the trivial and non--trivial sectors, and due to the triality property
of the $SO(8)$ representations. Thus, while in the new basis the appearance of
GUT symmetries is more cumbersome, it may be more instrumental in revealing
the duality symmetries that underly the string vacua.

\noindent

\end{abstract}
\smallskip}
\end{titlepage}

\setcounter{footnote}{0}

% ========================= DEFINITIONS ===================================
\def\beq{\begin{equation}}
\def\eeq{\end{equation}}
\def\beqn{\begin{eqnarray}}
\def\eeqn{\end{eqnarray}}

\def\no{\noindent }
\def\nolabel{\nonumber }
\def\ie{{\it i.e.}}
\def\eg{{\it e.g.}}
\def\half{{\textstyle{1\over 2}}}
\def\third{{\textstyle {1\over3}}}
\def\quarter{{\textstyle {1\over4}}}
\def\sixth{{\textstyle {1\over6}}}
\def\m{{\tt -}}
\def\p{{\tt +}}

\def\Tr{{\rm Tr}\, }
\def\tr{{\rm tr}\, }

\def\slash#1{#1\hskip-6pt/\hskip6pt}
\def\slk{\slash{k}}
\def\GeV{\,{\rm GeV}}
\def\TeV{\,{\rm TeV}}
\def\y{\,{\rm y}}
\def\SM{Standard--Model }
\def\SUSY{supersymmetry }
\def\SSSM{supersymmetric standard model}
\def\vev#1{\left\langle #1\right\rangle}
\def\l{\langle}
\def\r{\rangle}
\def\o#1{\frac{1}{#1}}

\def\Htw{{\tilde H}}
\def\chibar{{\overline{\chi}}}
\def\qbar{{\overline{q}}}
\def\ibar{{\overline{\imath}}}
\def\jbar{{\overline{\jmath}}}
\def\Hbar{{\overline{H}}}
\def\Qbar{{\overline{Q}}}
\def\abar{{\overline{a}}}
\def\alphabar{{\overline{\alpha}}}
\def\betabar{{\overline{\beta}}}
\def\tautwo{{ \tau_2 }}
\def\thetatwo{{ \vartheta_2 }}
\def\thetathree{{ \vartheta_3 }}
\def\thetafour{{ \vartheta_4 }}
\def\ttwo{{\vartheta_2}}
\def\tthree{{\vartheta_3}}
\def\tfour{{\vartheta_4}}
\def\ti{{\vartheta_i}}
\def\tj{{\vartheta_j}}
\def\tk{{\vartheta_k}}
\def\calF{{\cal F}}
\def\smallmatrix#1#2#3#4{{ {{#1}~{#2}\choose{#3}~{#4}} }}
\def\ab{{\alpha\beta}}
\def\Minv{{ (M^{-1}_\ab)_{ij} }}
\def\bone{{\bf 1}}
\def\ii{{(i)}}
\def\V{{\bf V}}
\def\N{{\bf N}}

% for basis vectors:
\def\b{{\bf b}}
\def\S{{\bf S}}
\def\X{{\bf X}}
\def\I{{\bf I}}
\def\mb{{\mathbf b}}
\def\mS{{\mathbf S}}
\def\mX{{\mathbf X}}
\def\mI{{\mathbf I}}
\def\balpha{{\mathbf \alpha}}
\def\bbeta{{\mathbf \beta}}
\def\bgamma{{\mathbf \gamma}}
\def\bxi{{\mathbf \xi}}

\def\t#1#2{{ \Theta\left\lbrack \matrix{ {#1}\cr {#2}\cr }\right\rbrack }}
\def\C#1#2{{ C\left\lbrack \matrix{ {#1}\cr {#2}\cr }\right\rbrack }}
\def\tp#1#2{{ \Theta'\left\lbrack \matrix{ {#1}\cr {#2}\cr }\right\rbrack }}
\def\tpp#1#2{{ \Theta''\left\lbrack \matrix{ {#1}\cr {#2}\cr }\right\rbrack }}
\def\l{\langle}
\def\r{\rangle}
\newcommand{\cc}[2]{c{#1\atopwithdelims[]#2}}
\newcommand{\nn}{\nonumber}

%================== BLACKBOARD BOLD CHARACTERS ==============================

\def\inbar{\,\vrule height1.5ex width.4pt depth0pt}

\def\IC{\relax\hbox{$\inbar\kern-.3em{\rm C}$}}
\def\IQ{\relax\hbox{$\inbar\kern-.3em{\rm Q}$}}
\def\IR{\relax{\rm I\kern-.18em R}}
 \font\cmss=cmss10 \font\cmsss=cmss10 at 7pt
\def\IZ{\relax\ifmmode\mathchoice
 {\hbox{\cmss Z\kern-.4em Z}}{\hbox{\cmss Z\kern-.4em Z}}
 {\lower.9pt\hbox{\cmsss Z\kern-.4em Z}}
 {\lower1.2pt\hbox{\cmsss Z\kern-.4em Z}}\else{\cmss Z\kern-.4em Z}\fi}

%========================================================================
%          MACROS FOR REFERENCES
%========================================================================
\def\AEF{A.E. Faraggi}
\def\JHEP#1#2#3{{\it JHEP}\/ {\bf #1} (#2) #3}
\def\NPB#1#2#3{{\it Nucl.\ Phys.}\/ {\bf B#1} (#2) #3}
\def\PLB#1#2#3{{\it Phys.\ Lett.}\/ {\bf B#1} (#2) #3}
\def\PRD#1#2#3{{\it Phys.\ Rev.}\/ {\bf D#1} (#2) #3}
\def\PRL#1#2#3{{\it Phys.\ Rev.\ Lett.}\/ {\bf #1} (#2) #3}
\def\PRT#1#2#3{{\it Phys.\ Rep.}\/ {\bf#1} (#2) #3}
\def\MODA#1#2#3{{\it Mod.\ Phys.\ Lett.}\/ {\bf A#1} (#2) #3}
\def\IJMP#1#2#3{{\it Int.\ J.\ Mod.\ Phys.}\/ {\bf A#1} (#2) #3}
\def\nuvc#1#2#3{{\it Nuovo Cimento}\/ {\bf #1A} (#2) #3}
\def\RPP#1#2#3{{\it Rept.\ Prog.\ Phys.}\/ {\bf #1} (#2) #3}
\def\EJP#1#2#3{{\it Eur.\ Phys.\ Jour.}\/ {\bf C#1} (#2) #3}
\def\etal{{\it et al\/}}

%==========================================================================
\hyphenation{su-per-sym-met-ric non-su-per-sym-met-ric}
\hyphenation{space-time-super-sym-met-ric}
\hyphenation{mod-u-lar mod-u-lar--in-var-i-ant}
%==========================================================================

%============================== SECTION 1 ============================

\setcounter{footnote}{0}
\section{Introduction}
String theory provides a unique phenomenological probe to
explore the unification of gravity and all other interactions including
gauge and Yukawa couplings.
String theory achieves this by providing a perturbatively
self--consistent calculational framework for quantum gravity,
while simultaneously giving rise to the gauge and matter structures
that are observed in high--energy experiments. Furthermore,
the gauge and matter sectors are imposed by the theory
self--consistency constraints. Given this unique status
a pivotal challenge is to construct string models that
reproduce the phenomenological subatomic data. In turn
such models are to be used to explore the properties of
string theory and its dynamics.

For over two decades the free fermionic construction of the
heterotic string \cite{fff1,fff2} provided the tools to develop
phenomenological string models \cite{ffm}. Three generation models
with the correct
Standard Model charge assignments, as well as the canonical $SO(10)$
embedding of the weak hypercharge, were constructed. Various issues
pertaining to the phenomenological Standard Model data and grand unification
were further explored in the framework of these models.

The existence of quasi--realistic free fermionic constructions justifies
the effort to better understand the properties of these models and the
global structures that underly them. In the orbifold language the
free fermionic construction
correspond to symmetric, asymmetric or freely acting
orbifolds. A subclass of them correspond to
symmetric $Z_2\times Z_2$ orbifold compactifications at
enhanced symmetry points in the toroidal moduli space \cite{z2z21,z2z22}.
Also the chiral matter spectrum arises from
twisted sectors and thus does not depend on the moduli. This
facilitates the complete classification of the topological sectors of the
$Z_2\times Z_2$
symmetric orbifolds. For type II string $N=2$ supersymmetric vacua
the general free fermionic classification techniques were developed in ref.
\cite{gkr}. The method was extended in refs.
\cite{fknr,nooij,fkr} for the classification of heterotic $Z_2\times Z_2$
orbifolds. In this class of models the six dimensional internal manifold
contains three twisted sectors.
In the heterotic string each of these sectors
may, or may not, a priori
(prior to application of the Generalised GSO (GGSO) projections),
give rise to spinorial representations.

The classification of heterotic $N=1$ vacua revealed a symmetry
in the distribution of $Z_2\times Z_2$ string vacua
under exchange of vectorial, and spinorial plus anti--spinorial,
representations of $SO(10)$ \cite{fkr},
which is akin to mirror symmetry \cite{mirror, vafawitten}
that exchanges spinorial with anti--spinorial representations.
The symmetry under the exchange of spinorial plus anti--spinorial
representations with vectorial representations
is evident when the $SO(10)$ symmetry is
enhanced to $E_6$, in which case $\# (16+{\overline 16}) = \# (10)$.
We demonstrated in ref. \cite{fkr} that the symmetry persists also when
there is no enhancement to $E_6$, and
the existence of self--dual vacua in which
$\# (16+{\overline 16}) = \# (10)$, but in which the $SO(10)$ symmetry
is not enhanced to $E_6$.

The existence of the spinor--vector duality over the entire class
of symmetric $Z_2\times Z_2$ orbifolds indicates a global structure
that underlies this entire space of vacua.
It was noted in ref. \cite{fkr}
that the symmetry operates separately
on each of the three twisted sectors of the $Z_2\times Z_2$ orbifold.
Since each of the twisted sectors of the $Z_2\times Z_2$ orbifold preserves
$N=2$ space--time supersymmetry, the spinor--vector duality should
already exist at the level of $N=2$ vacua. That is it should exist
also in models in which the $N=2$ space--time supersymmetry is not
broken to $N=1$. This fact is an important clue in trying to understand
the origin of the spinor--vector duality and the global structures
that underly the free fermionic models, as well as
the $Z_2\times Z_2$ orbifold constructions.

In this paper we show the existence of the spinor--vector
duality in $N=2$ vacua. This is demonstrated by generating the
complete space of $N=2$ vacua, as well as by presenting an
algebraic proof of the duality map. In the first instance
the $N=2$ models can be generated by removing from the basis set
of ref. \cite{fkr} the basis vector that breaks $N=2$ space--time
supersymmetry to $N=1$.
To further elucidate the existence of the duality symmetry
we will use for our construction
a new basis to generate the space of free fermionic
$Z_2\times Z_2$ orbifolds. In the new basis the untwisted gauge
symmetry is reduced to $SO(12)\times SO(8)^3\times SO(2)^4$.
In the new basis the GUT $SO(10)$ symmetry is obtained
by enhancement of an $SO(8)\times SO(2)$ untwisted group factor with
additional vector bosons from non--trivial sectors. Thus,
the existence of a GUT symmetry is obscured in this new basis.
On the other hand the existence of a map between
spinors and vectors becomes more transparent, as it is
generated by the $U(1)$ current of a ``would--be
$N=2$ world--sheet supersymmetry'' in the non--supersymmetric
side of the heterotic--string.

Our paper is organised as follows: in section \ref{neq2}
we discuss the method of classification of the $N=2$ space--time
supersymmetric vacua.
In section \ref{proof} we present an algebraic proof of the spinor
vector--duality in the case of $N=2$ free fermionic vacua. In section
\ref{newbasis} we present a new basis to generate the space of
free fermionic vacua. In the new basis the GUT symmetries are
generated from trivial and non--trivial sectors. The primary feature
of the new basis is the division of the gauge degrees of freedom
of the heterotic string into four blocks of $SO(8)$ characters. Thus,
while the origin of the GUT symmetries is obscured, the duality properties
of the heterotic string vacua are more transparent in the new basis.
Section \ref{conclude} concludes the paper.

\section{\bf $N=2$ model classification}\label{neq2}

In the free fermionic formulation the 4-dimensional heterotic string,
in the light-cone gauge, is described
by $20$ left--moving  and $44$ right--moving two dimensional real
fermions \cite{fff1,fff2}.
A large number of models can be constructed by choosing
different phases picked up by   fermions ($f_A, A=1,\dots,44$) when transported
along
the torus non-contractible loops.
Each model corresponds to a particular choice of fermion phases consistent with
modular invariance
that can be generated by a set of  basis vectors $v_i,i=1,\dots,n$,
$$v_i=\left\{\alpha_i(f_1),\alpha_i(f_{2}),\alpha_i(f_{3}))\dots\right\}$$
describing the transformation  properties of each fermion
\begin{equation}
f_A\to -e^{i\pi\alpha_i(f_A)}\ f_A, \ , A=1,\dots,44~.
\end{equation}
The basis vectors span a space $\Xi$ which consists of $2^N$ sectors that give
rise to the string spectrum. Each sector is given by
\begin{equation}
\xi = \sum N_i v_i,\ \  N_i =0,1
\end{equation}
The spectrum is truncated by a GGSO projection whose action on a
string state  $|S>$ is
\begin{equation}\label{eq:gso}
e^{i\pi v_i\cdot F_S} |S> = \delta_{S}\ \cc{S}{v_i} |S>,
\end{equation}
where $F_S$ is the fermion number operator and $\delta_{S}=\pm1$ is the
spacetime spin statistics index.
Different sets of projection coefficients $\cc{S}{v_i}=\pm1$ consistent with
modular invariance give
rise to different models. Summarising: a model can be defined uniquely by a set
of basis vectors $v_i,i=1,\dots,n$
and a set of $2^{N(N-1)/2}$ independent projections coefficients
$\cc{v_i}{v_j}, i>j$.

The two dimensional
free fermions in the light-cone gauge (in the usual notation
\cite{fff1,fff2,ffm}) are:
$\psi^\mu, \chi^i,y^i, \omega^i, i=1,\dots,6$ (real left-moving fermions)
and
$\bar{y}^i,\bar{\omega}^i, i=1,\dots,6$ (real right-moving fermions),
${\bar\psi}^A, A=1,\dots,5$, $\bar{\eta}^B, B=1,2,3$, $\bar{\phi}^\alpha,
\alpha=1,\ldots,8$ (complex right-moving fermions).
The class of models under investigation,
is generated by a set $V$ of 11 basis vectors
$$
V=\{v_1,v_2,\dots,v_{11}\},
$$
where
\begin{eqnarray}
v_1=1&=&\{\psi^\mu,\
\chi^{1,\dots,6},y^{1,\dots,6}, \omega^{1,\dots,6}| \nn\\
& & ~~~\bar{y}^{1,\dots,6},\bar{\omega}^{1,\dots,6},
\bar{\eta}^{1,2,3},
\bar{\psi}^{1,\dots,5},\bar{\phi}^{1,\dots,8}\},\nn\\
v_2=S&=&\{\psi^\mu,\chi^{1,\dots,6}\},\nn\\
v_{2+i}=e_i&=&\{y^{i},\omega^{i}|\bar{y}^i,\bar{\omega}^i\}, \
i=1,\dots,6,\nn\\
v_{9}=b_1&=&\{\chi^{34},\chi^{56},y^{34},y^{56}|\bar{y}^{34},
\bar{y}^{56},\bar{\eta}^1,\bar{\psi}^{1,\dots,5}\},\label{basis}\\
v_{10}=z_1&=&\{\bar{\phi}^{1,\dots,4}\},\nn\\
v_{11}=z_2&=&\{\bar{\phi}^{5,\dots,8}\}.\nn
\end{eqnarray}
The minimal gauge group is
$$
SO(12)\times{SU(2)}_1\times{SU(2)}_2\times{SO(8)}_1\times{SO(8)}_2
$$
Various extensions are possible since extra massless states can
arise from
$x,z_1,z_2,z_1+z_2$,
where the anti--holomorphic $x$ set is
\ba
x=1+S+\sum_{i=1}^6e_i+\sum_{k=1}^2 z_k
=\{{\bar{\eta}^{123},\bar{\psi}^{12345}}\}\ .
\ea
Among these massless states there are also space--time vector
bosons, which extend the four dimensional gauge symmetry group, possibly
also mixing the observable and hidden sectors gauge groups.
As we discuss further below
a choice GGSO projection coefficients exists which avoids such mixings.

Spinorial representations of the $SO(12)$ GUT group are
in the ({\bf 32},{\bf1},{\bf1}),
({\bf 32$^\prime$},{\bf1},{\bf1}) of the
$SO(12)\times{SU(2)}_1\times{SU(2)}_2$ observable gauge group.
These representations arise from the twisted sector
\begin{eqnarray}
B_{p^Sq^Sr^Ss^S}&=&S+b_1+p^S e_3+q^S e_4 +
r^S e_5 + s^S e_6
\end{eqnarray}
where $p^S,q^S,r^S,s^S=\{0,1\}$.
In this sectors the six complex world--sheet fermion
$\{{\bar\psi}^{1,\cdots,5},{\bar\eta}^1\}$ are periodic, and there
are no oscillators acting on the non--degenerate vacuum in this sector.
Spinorial representations of the hidden ${SO(8)}_i, (i=1,2),$
arise from the sectors
\ba
H^i_{k^S\ell^Sm^Sn^S} &=& \nonumber\\
& & S+b_1+x+z_i+k_i^S e_3+\ell_i^S e_4 +
m_i^S e_5 + n_i^S e_6\ ,\ i=1,2~~~
\ea
In these sectors the corresponding $\{{\bar\phi}^{1,\dots,4}\}$
or $\{{\bar\phi}^{5,\dots,8}\}$ are periodic and again there are
no oscillators acting on the non--degenerate vacuum in these sectors.
States in the vectorial representations of the $SO(12)$ GUT group,
{\it i.e.} in the ({\bf 12},{\bf2},{\bf1}) and ({\bf 12},{\bf1},{\bf2}),
of the observable $SO(12)\times SU(2)\times SU(2)$ gauge group,
as well as states in the vectorial representations of the hidden
$SO(8)_i$ gauge groups  arise from the sector
%Vectorial representations of the observab%le, as well as ve
%Observable  sector vectorials
%{\it i.e.}
%
%({\bf 1},{\bf2},{\bf1}),
%({\bf 1},{\bf1},{\bf2}) hidden
%sector vectorials  arise from
\begin{eqnarray}
& & V_{p^Vq^Vr^Vs^V} ~=~
  B_{p^Vq^Vr^Vs^V}+x  \nonumber\\
 = & & S+b_1+x+p^V e_3+q^V e_4 + r^V e_5 + s^V e_6~~~
\end{eqnarray}
in this sector the world--sheet complex fermions $\{{\bar\eta}^{2,3}\}$
are periodic. The massless states are obtained by acting
with a fermionic oscillator on the non--degenerate vacuum.
Following the methodology of ref. \cite{fkr} the
GGSO projections are translated to a set of algebraic equations.
The number of observable spinorials $S$ and vectorials $V$,
as well as the number of hidden sector spinorials $S_1, S_2$
and vectorials
$V_1,V_2$ are determined by the solutions of the equations
\ba
\Delta\,~~U_S &=& Y_S\ , \nonumber \\
\Delta\,~~U_V &=& Y_V\ , \nonumber \\
\Delta_i\,~~U_{S}^i & =& Y_{S}^i\ ,\ i=1,2\ , \nonumber\\
\Delta\,~U_{V}^i & =& Y_{V}^i\ , \ i=1,2\ ,
\label{deltaUeqY}
\ea
where the unknowns are the fixed point labels
\ba U_{S}= \left[
\begin{array}{c}
p^{S}\\
q^{S}\\
r^{S}\\
s^{S}
\end{array}
\right]
\ ,\
U_{V}= \left[
\begin{array}{c}
p^{V}\\
q^{V}\\
r^{V}\\
s^{V}
\end{array}
\right]
\ ,\
U^i_{S}= \left[
\begin{array}{c}
k^{S}_i\\
\ell^{S}_i\\
m^{S}_i\\
n^{S}_i
\end{array}
\right]
\ ,\
U^i_{V}= \left[
\begin{array}{c}
k^{V}_i\\
\ell^{V}_i\\
m^{V}_i\\
n^{V}_i
\end{array}
\right]
\ .
\label{ucoef}
%U_{1}= \left[
%\begin{array}{c}
%p^{1}\\
%q^{1}\\
%r^{1}\\
%s^{1}
%\end{array}
%\right]
%\ ,\
%U_{1'}= \left[
%\begin{array}{c}
%p^{1'}\\
%q^{1'}\\
%r^{1'}\\
%s^{1'}
%\end{array}
%\right]
\ea
In what follows it is convenient to introduce the phases
$\oo{a_i}{a_j}$, which are defined via the GGSO projection
coefficients as
\ba
\cc{a_i}{a_j}=e^{i \pi \oo{a_i}{a_j}}\,\ ,\  \oo{a_i}{a_j}=0,1
\ea
with the properties
\ba
\oo{a_i}{a_j+a_k}&=&\oo{a_i}{a_j}+\oo{a_i}{a_k} \ ,
\forall\ a_{i}: \{\psi^\mu\}\cap a_i=\O
\\
\oo{a_i}{a_j}&=&\oo{a_j}{a_i} \ ,
\forall\ a_{i},a_{j}: a_i\cdot a_j=0\ {\rm mod}\ 4
\ea
where $\# (a_i\cdot a_j)\equiv\# \left[a_i\cup a_j - a_i\cap a_j\right]$.
On the left--hand side of the algebraic GGSO equations (\ref{deltaUeqY})
the $\Delta$ operators are binary matrices composed of
the relevant GGSO phases.
\ba
\Delta&=&\left[
\begin{array}{cccc}
\oo{e_1}{e_3}&\oo{e_1}{e_4}&\oo{e_1}{e_5}&\oo{e_1}{e_6}\\
\oo{e_2}{e_3}&\oo{e_2}{e_4}&\oo{e_2}{e_5}&\oo{e_2}{e_6}\\
\oo{z_1}{e_3}&\oo{z_1}{e_4}&\oo{z_1}{e_5}&\oo{z_1}{e_6}\\
\oo{z_2}{e_3}&\oo{z_2}{e_4}&\oo{z_2}{e_5}&\oo{z_2}{e_6}
\end{array}
\right]
 \nn\\
 \Delta_1&=&\left[
\begin{array}{cccc}
\oo{e_1}{e_3}&\oo{e_1}{e_4}&\oo{e_1}{e_5}&\oo{e_1}{e_6}\\
\oo{e_2}{e_3}&\oo{e_2}{e_4}&\oo{e_2}{e_5}&\oo{e_2}{e_6}\\
\oo{z_2}{e_3}&\oo{z_2}{e_4}&\oo{z_2}{e_5}&\oo{z_2}{e_6}
\end{array}
\right]\nn\\
\Delta_2&=&\left[
\begin{array}{cccc}
\oo{e_1}{e_3}&\oo{e_1}{e_4}&\oo{e_1}{e_5}&\oo{e_1}{e_6}\\
\oo{e_2}{e_3}&\oo{e_2}{e_4}&\oo{e_2}{e_5}&\oo{e_2}{e_6}\\
\oo{z_1}{e_3}&\oo{z_1}{e_4}&\oo{z_1}{e_5}&\oo{z_1}{e_6}
\end{array}
\right]\ , \label{deltas}
\ea
whereas the right--hand sides of the GGSO projection equations
are composed of one column vectors appropriate for the respective
sectors,
\ba
Y_{S}=
\left[
\begin{array}{c}
\oo{e_1}{b_1}\\
\oo{e_2}{b_1}\\
\oo{z_1}{b_1}\\
\oo{z_2}{b_1}
\end{array}
\right]\ &,& \
Y_{V}=
\left[
\begin{array}{c}
\oo{e_1}{b_1+x}\\
\oo{e_2}{b_1+x}\\
\oo{z_1}{b_1+x}\\
\oo{z_2}{b_1+x}
\end{array}
\right]
\nonumber\\
Y_{S'}^{(1)}=
\left[
\begin{array}{c}
\oo{e_1}{b_1+x+z_1}\\
\oo{e_2}{b_1+x+z_1}\\
\oo{z_2}{b_1+x+z_1}
\end{array}
\right]
\ &,&\
Y_{V'}^{(1)}=
\left[
\begin{array}{c}
\oo{e_1}{b_1+x}\\
\oo{e_2}{b_1+x}\\
\oo{z_1}{b_1+x}+1\\
\oo{z_2}{b_1+x}
\end{array}
\right]
\nonumber\\
Y_{S'}^{(2)}=
\left[
\begin{array}{c}
\oo{e_1}{b_1+x+z_2}\\
\oo{e_2}{b_1+x+z_2}\\
\oo{z_1}{b_1+x+z_2}
\end{array}
\right]
\ &,& \
Y_{V'}^{(2)}=
\left[
\begin{array}{c}
\oo{e_1}{b_1+x}\\
\oo{e_2}{b_1+x}\\
\oo{z_1}{b_1+x}\\
\oo{z_2}{b_1+x}+1
\end{array}
\right]\ . \label{yvectors}
%Y_{1}=
%\left[
%\begin{array}{c}
%\oo{e_1}{b_1+x}+1\\
%\oo{e_2}{b_1+x}\\
%\oo{z_2}{b_1+x}
%\end{array}
%\right]
%\ &,&\
%Y_{1'}=
%\left[
%\begin{array}{c}
%\oo{e_1}{b_1+x}\\
%\oo{e_2}{b_1+x}+1\\
%\oo{z_1}{b_1+x}
%\end{array}
%\right]\ .
\ea
We note that the $\Delta$ matrices of the observable $SO(12)$
spinorial and vectorial representations are identical, and that the
two column vectors $Y_S$ and $Y_V$ are ``mapped'' by the addition of the
vector $x$.
Following the methods developed in \cite{fkr}
the number of $N=2$ hypermultiplets in the
$SO(12)$ spinorial ($S$) and vectorial ($V$) representations
are given by
\ba
S=\left\{
\begin{array}{ll}
2^{4-{\rm rank}\left(\Delta\right)}
&,\ {\rm rank}\left(\Delta\right)=
{\rm rank}\left[\Delta,Y_{S}\right]\\
 & \\
0&,\ {\rm rank}\left(\Delta\right)<{\rm rank}
\left[\Delta,Y_{S}\right]
\end{array}
\right.
\label{ssi}
\ea
\ba
V=\left\{
\begin{array}{ll}
2^{4-{\rm rank}\left(\Delta\right)}&,\ {\rm rank}
\left(\Delta\right)={\rm
rank}\left[\Delta,Y_{V}\right]\\
 & \\
0&,\ {\rm rank}\left(\Delta\right)<{\rm rank}
\left[\Delta,Y_{V}\right]
\end{array}
\right.
\label{vvi}
\ea
where the respective
$\left[\Delta,Y\right]$ are the augmented matrices.
Similar results hold for the counting of
${SO(8)}_k\,,k=1,2$ representations.

\subsection{The four dimensional gauge group}
For all the models generated by the basis set (\ref{basis})
gauge bosons arise from the following four sectors :
$$G=\{0,z_1,z_2,z_1+z_2,x\}$$

The null sector gauge bosons give rise to the gauge symmetry
\begin{equation}
U(1)^2\times SO(12)\times {SU(2)}^2 \times SO(8)^2.
\label{mg}
\end{equation}

The first two $U(1)'s$ arise from the world--sheet complexified
fermions $\zeta^i=1/\sqrt{2}({\bar y}^i+i{\bar\omega}^i)$,
$(i=1,2)$, whereas the two $SU(2)'s$ arise from the
complex world--sheet fermions ${\bar\eta}^i$ $(i=2,3)$.
The remaining group factors arise from the world--sheet fermions
$\{{\bar\psi}^{1,\cdots, 5}, {\bar\eta}^1\}$, ${\bar\phi}^{1,\cdots,4}$
and ${\bar\phi}^{5,\cdots,8}$, respectively.

The $x$ gauge bosons when present lead to enhancements of the
$SO(12)$) gauge group, while
the $z_1+z_2$ sector can enhance the hidden sector ($SO(8)^2$).
The $z_1, z_2$ sectors accept oscillators that can
also give rise to mixed type gauge bosons and completely reorganise
the gauge group.
The appearance of mixed states is in general controlled by the
phase $\cc{z_1}{z_2}$. The choice $\cc{z_1}{z_2}=+1$ allows for
mixed gauge bosons
and leads to the gauge groups presented in Table \ref{gga}.
\begin{table}
\centering
\begin{tabular}{|c|c|c|c|c|c|c|c|c|}
\hline
$\cc{z_1}{z_2}$&$\cc{b_1}{z_1}$&$\cc{b_2}{z_1}$&
$\cc{b_1}{z_2}$&$\cc{b_2}{z_2}$&$\cc{e_1}{z_1}$&$\cc{e_2}{z_2}$
&$\cc{e_1}{e_2}$&Gauge group\\
\hline
$+$&$+$&$+$&$+$&$+$&$+$&$+$&$+$&
$U(1)^2 \times SO(12)\times{SO(18)}$\\
\hline
$+$&$+$&$+$&$+$&$+$&$-$&$-$&$+$&
$SO(12)\times{SO(4)}\times{SO(10)^2}$\\
\hline
$+$&$+$&$+$&$+$&$+$&$-$&$+$&$+$&
$U(1)\times SO(12)^2\times{SO(10)}$\\
\hline
$+$&$-$&$-$&$-$&$-$&$+$&$+$&$+$&
$U(1)^2\times SO(28)\times{SO(4)}$\\
\hline
$-$&$+$&$+$&$+$&$+$&$+$&$+$&$+$&
$U(1)^2\times E_7\times{SU(2)}\times{E_8}$\\
\hline
$-$&$-$&$+$&$+$&$-$&$+$&$+$&$+$&
$U(1)^2\times E_7\times{SU(2)}\times{SO(16)}$\\
\hline
$-$&$+$&$+$&$+$&$+$&$+$&$+$&$-$&
$U(1)^2\times SO(12)\times{SO(4)}\times{E_8}$\\
\hline
$-$&$+$&$+$&$+$&$+$&$-$&$-$&$-$&
$U(1)^2\times SO(12)\times{SO(4)}\times{SO(8)}^2$\\
\hline
\end{tabular}
\caption{\label{gga}\it Typical enhanced gauge groups  and
associated projection coefficients for a generic model generated
by the basis (\ref{basis})(coefficients not included
equal to +1 except those fixed by space-time supersymmetry
and conventions).
}
\end{table}

The choice  $\cc{z_1}{z_2}=-1$ eliminates all mixed gauge
bosons and there are a few possible enhancements:
${SO(12)}\times{SU(2)}\to{E_7}$ and/or ${SO(8)}^2\to\{SO(16),E_8\}$.
The additional gauge bosons that may arise from the sectors
$$
x\ ,\ z_1\ ,\ z_2\ ,\ z_1+z_2
$$
can lead to enhancements of the observable and/or the hidden gauge group.
These enhancements are model dependent, and hence depend on specific choices
of GGSO phases. These enhancements include:
\newcounter{mlls}
\begin{list}{(\Roman{mlls})}{\usecounter{mlls}}
\item The $x$--sector gauge bosons give rise to
$SO(12)\times{SU(2)}\to{E_7}$ enhancement when
\ba
\oo{e_i}{x}=\oo{z_k}{x}=0\ \forall\ i=1,\dots,6\ ,\ k=1,2~.
\label{eixr}
\ea
\item The $z_1+z_2$--sector
gauge bosons can lead to ${SO(8)}^2\to{SO(16)}$
enhancement when
\ba
\oo{e_i}{z_1}=\oo{e_i}{z_2}\ \forall\ i=1,\dots,6\ ,\
\oo{b_1}{z_1}=\oo{b_1}{z_2}\
\label{eiz1bmz1}
\ea
\item The $z_k$--sectors, $(k=1,2)$,
enhancements involve right--moving fermionic
oscillators and belong in two classes depending on the
value of $\oo{z_1}{z_2}$:

(a) for $\oo{z_1}{z_2}=1$ we obtain gauge bosons that
involve $z_1$ and/or $z_2$ oscillators, namely
$\{\bar{\phi}^{1\dots8}\}$. These lead
to hidden group enhancements, and particularly to
${SO(8)}^2\to{SO(16)}$ when
\ba
\oo{z_1}{z_2}=1\ ,\ \oo{e_i}{z_k}=\oo{b_1}{z_k}=0\ \forall\ i=1,\dots,6
\label{z1z2eizk}
\ea
(b) for $\oo{z_1}{z_2}=0$ we obtain gauge bosons that involve
oscillators not included in $z_1,z_2$ and lead
thus to gauge bosons that mix
${SO(8)}_1$ or/and  ${SO(8)}_2$  with other  group factors in
(\ref{mg}).
These include:

The case $\oo{z_k}{b_1}=1$ selects vector bosons that enhance
the $SO(12)\times SO(8)_k$ depending on the choices
of $\oo{z_1}{e_i}$ $(i=1,\cdots, 6)$. The basis vectors
$e_{1,2}$ acts as projectors on these states. Setting
$\oo{z_k}{e_1}=\oo{z_k}{e_2}=0$ keeps the states in the spectrum,
whereas $\oo{z_k}{e_1}=1$ and/or $\oo{z_k}{e_2}=1$ projects them
out. The remaining
$\oo{z_k}{e_i}$ phases select particular states according to:
\beqn
\oo{z_k}{e_{3,4,5,6}} ~~~~~= 0  ~~~~~~~~~~~~~~~~~~~~~~~~~~~~~~~~~~~~&
                \rightarrow (12,8_k)\label{e1234}\\
\oo{z_k}{e_{i}}  ~~~~~~~~~~= 1 ~~\&~\oo{z_k}{e_{j,k,l}}=0 ~~~~~~~~~~~~&
                \rightarrow (~1,8_k)\label{ei}\\
\oo{z_k}{e_{i,j}} ~~~~~~~~~= 1 ~~\&~\oo{z_k}{e_{k,l}}~=0 ~~~~~~~~~~~~&
                \rightarrow (~1,1_k)\label{eij}\\
\oo{z_k}{e_{i,j,k}} ~~~~~~~= 1 ~~\&~\oo{z_k}{e_{l}}~~~=0 ~~~~~~~~~~~~&
                \rightarrow (~1,1_k)\label{eijk}\\
\oo{z_k}{e_{i,j,k,l}}~~~~~~= 1 ~~~~~~~~~~~~~~~~~~~~~~~~~~~~~~~~~~~&
                \rightarrow (~1,1_k)\label{eijkl}\\
 {\rm with}~
\{i\ne j\ne k\ne l\}~=~\{3,4,5,6\}& \nonumber\\
\eeqn
Case (\ref{e1234}) enhances the $SO(12)\times SO(8)$ symmetry to $SO(20)$.
Case (\ref{ei})    enhances the $SO(12)\times SO(8)$ symmetry to
                            $SO(12)\times SO(9)$.
Cases (\ref{eij}), (\ref{eijk})
and (\ref{eijkl}) project the additional vector
bosons from the sectors $z_k$, and leave the $SO(12)\times SO(8)$ symmetry
unenhanced.

The case $\oo{z_k}{b_1}=0$ selects vector bosons that enhance
the $U(1)^2\times SO(4)\times SO(8)_k$ symmetry depending on the choices
of $\oo{z_k}{e_i}$ $(i=1,\cdots, 6)$.
The $\oo{z_k}{e_i}$ phases select particular states according to:
\beqn
\oo{z_k}{e_{1,2,3,4,5,6}} = 0 ~~~~~~~~~~~~~~~~~~~~~~~~~~~~~~~~~~~~
            \rightarrow (0^2,4,8_k)~~~~~~& \label{b1pe1234}\\
\oo{z_k}{e_{i}} ~~~~~~~~= 1 ~\&~\oo{z_k}{e_{j,k,l,m,n}}=0 ~~~~~~~~~
            \rightarrow (\pm1_i,0_j,1,8_k)& \label{b1pei}\\
 {\rm with}~
\{i\ne j\} = 1 {\rm~or}~2 \ne \{k\ne l\ne m\ne n\}~=~\{3,4,5,6\}
~~~~~~~~& \nonumber\\
\oo{z_k}{e_{i}} ~~~~~~~~= 1 ~\&~\oo{z_k}{e_{j,k,l,m,n}}=0 ~~~~~~~~~
            \rightarrow (0^2,1,8_k) ~~~~~~& \label{b1pei3456}\\
 {\rm with}~
\{i\} = \{3,4,5,6\} \ne \{j\ne k\ne l\ne m\ne n\}~=~\{1,2,3,4,5,6\}
& \nonumber\\
\oo{z_k}{e_{i,j}}~~~~~~~= 1 ~\&~\oo{z_k}{e_{k,l}}=0 ~~~~~~~~~~~~~~~
            \rightarrow (0^2,1,1_k) ~~~~~~& \label{b1peij}\\
\oo{z_k}{e_{i,j,k}} ~~~~~~=1 ~\&~\oo{z_k}{e_{l,m,n}}=0 ~~~~~~~~~~~~
            \rightarrow (0^2,1,1_k) ~~~~~~& \label{b1peijk}\\
\oo{z_k}{e_{i,j,k,l}} ~~~~~=1 ~\&~\oo{z_k}{e_{m,n}}=0 ~~~~~~~~~~~~
            \rightarrow (0^2,1,1_k) ~~~~~~& \label{b1peijkl}\\
\oo{z_k}{e_{i,j,k,l,m}} ~~~=1 ~\&~\oo{z_k}{e_{l,m,n}}=0 ~~~~~~~~~~~
            \rightarrow (0^2,1,1_k) ~~~~~~& \label{b1peijklm}\\
\oo{z_k}{e_{i,j,k,l,m,n}} ~=1 ~~~~~~~~~~~~~~~~~~~~~~~~~~~~~~~~~~~
            \rightarrow (0^2,1,1_k) ~~~~~~& \label{b1peijklmn}\\
 {\rm with}~
\{i\ne j\ne k\ne l\ne m\ne n\}~=~\{1,2,3,4,5,6\}~~~~~~~~~~~~~~~~~~~& \nonumber
\eeqn
for $k=1$ or/and $k=2$. In this case the gauge group enhancement
includes several possibilities,
depending on the $\oo{b_1}{z_k}$ we can obtain:\\

Case (\ref{b1pe1234}) enhances the $U(1)^2\times SO(4)\times SO(8)_k$
symmetry to $U(1)^2\times SO(12)$.
Case (\ref{b1pei})    enhances the
$U(1)_i\times U(1)_j\times SO(4)\times SO(8)_k$ symmetry to
                $U(1)_j\times SO(4)\times SO(10)$.
Case (\ref{b1pei3456}) enhances the $SO(8)_k$ symmetry to $SO(9)_k$.
Cases (\ref{b1peij}), (\ref{b1peijk}), (\ref{b1peijkl}), (\ref{b1peijklm})
and
(\ref{b1peijklmn}) project the additional vector
bosons from the sectors $z_k$, and leave the
$U(1)^2\times SO(4)\times SO(8)_k$ symmetry unenhanced.

Depending on the separate enhancements of $SO(8)_k$ for $k=1,2$ we can
obtain for example:

$SO(12)\times{SO(8)}_k\to{SO(20)}$,\\
$SO(4)\times{SO(8)}_k\to{SO(12)}$,\\
${SO(8)}_k\times{U(1)}\to{SO(10)}_k$,\\
or
${SO(8)}^2\times{U(1)}^2\to{SO(10)}^2$.

Moreover for $\oo{z_1}{z_2}=0$ and particular choices
of $\oo{e_i}{z_k}$ and $\oo{b_1}{z_k}$ we can have
${SO(8)}_k\to{SO(9)}$ enhancements.\\
\end{list}
Mixed combinations of the above are possible when the
conditions on the associated GGSO coefficients are
compatible. For example combination of gauge bosons (II)
with those in (IIIb) can lead to
$SO(12)\times{SO(8)}^2\to{SO(28)}$ enhancement.

In the present work
we restrict to models where all the additional gauge bosons from the
sectors $x$, $z_1+z_2$ and $z_k$ sectors
are absent.
This is achieved for appropriate choice of the GGSO phases
such that the above requirements
%(\ref{eixr}--\ref{z1z2eizk2})
are not satisfied.

\section{Results}

\begin{figure}[!ht]
\centering
\includegraphics{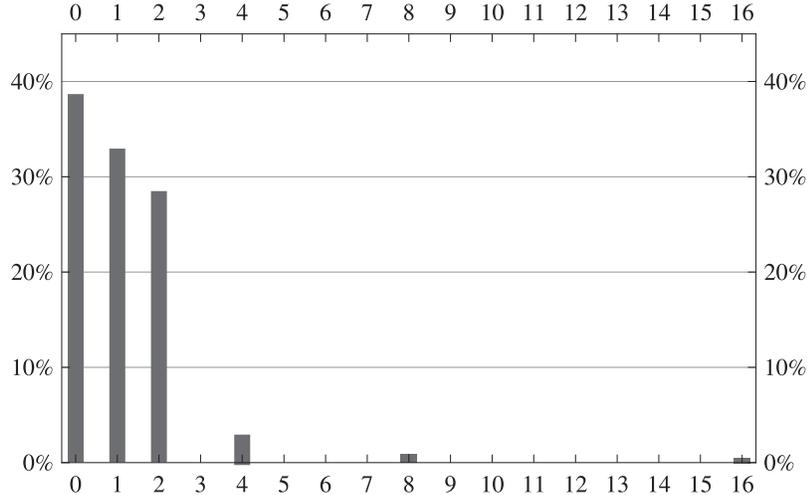}
\caption{\label{sa}\it Percentage of models versus the number of
$N=2$ $SO(12)$ spinorial/vectorial multiplets.}
\end{figure}

Using the results of section \ref{neq2}, 
we can calculate the number of $SO(12)$ spinorials ($S$) and vectorials
($V$) as well as the numbers of  ${SO(8)}_k, k=1,2$ spinors and
vectors for a given set of GSO projection coefficients.
They turn out to depend on 26 parameters, namely  
$\oo{e_1}{e_j}, \oo{e_2}{e_j}, \oo{z_1}{e_j}, \oo{z_2}{e_j}\, j=3,\dots,6 $,
$\oo{e_1}{e_2}, \oo{e_1}{z_k}, \oo{e_2}{z_k}, \oo{e_k}{b_1},
\oo{z_k}{b_1}\,, k=1,2$ and $\oo{z_1}{z_2}$
giving rise to $2^{26}$ distinct models.
The full set of models can be classified with the help of a computer programme
following the methods developed in \cite{fkr}.
As far as the total number of twisted $SO(12)$ spinorials or
vectorials is concerned
we find that only models with $S,V=0,1,2,4,8,16$ are allowed.
A graphical representation
of the percentage of distinct models versus the number
of $SO(12)$ spinorials/vectorials is presented in Figure \ref{sa}.
These results were obtained by a Monte--Carlo analysis that generates
random choices of the GGSO phases. In this sense the results shown in
fig. \ref{sa} are based on a statistical polling.
We note that analysis of large sets of string vacua has also been carried
out by other groups \cite{statistical}.

\section{Analytic Proof of Spinor-Vector Duality}\label{proof}
As seen from eqs. (\ref{deltaUeqY}--\ref{yvectors})
the number of $SO(12)$ vectorial and spinorial representations
are interchanged when the ranks of the associated $Y$-vectors are
interchanged ($Y_S\leftrightarrow Y_V$)
\ba
{\rm rank}\left[\Delta,Y_{S}\right]\leftrightarrow{\rm rank}
\left[\Delta,Y_{V}\right]~,
\ea
as follows from eqs. (\ref{ssi}) and (\ref{vvi}).
In order to prove the existence of $V\leftrightarrow S_t$ duality
we have to demonstrate the existence of a universal map that preserves
the ranks of the matrices, while exchanging
$Y_S\leftrightarrow Y_V$. Since the rank of
the augmented matrix does not change by adding to the
$Y_V$ the sum of the columns of $\Delta$
we can rewrite $Y_V$ as follows

\begin{align}
Y_V
=
\left[
\begin{array}{c}
\oo{e_1}{b_1}+\oo{e_1}{e_2+z_1+z_2}\\
\oo{e_2}{b_1}+\oo{e_2}{e_1+z_1+z_2}\\
\oo{z_1}{b_1}+\oo{z_1}{e_1+e_2+z_2}+1\\
\oo{z_2}{b_1}+\oo{z_2}{e_1+e_2+z_1}+1
\end{array}
\right]
=
Y_S+
\left[
\begin{array}{c}
\oo{e_1}{e_2+z_1+z_2}\\
\oo{e_2}{e_1+z_1+z_2}\\
\oo{z_1}{e_1+e_2+z_2}+1\\
\oo{z_2}{e_1+e_2+z_1}+1
\end{array}
\right]
\end{align}
The last vector in the above equation contains six independent phases that do not appear in $\Delta$
namely $\oo{e_1}{e_2},\oo{e_1}{z_1},\oo{e_1}{z_2}, \oo{e_2}{z_1},\oo{e_2}{z_2},\oo{z_1}{z_2}$. Four of them
can always be used to realize the $V\leftrightarrow S$ exchange.

\section{A novel basis}\label{newbasis}
We present a new basis for generating the free fermionic models that
may shed new light on the structure of heterotic--string unification,
on  its relation to the low energy data
and to other limits of string theory.
The new basis is obtained by splitting the gauge degrees of freedom
of the uncompactified ten dimensional theory
into four equivalent subgroups. In effect, this entails that the untwisted
vector bosons produce the generators of $SO(8)^4$ gauge group.
This is achieved by introducing the four basis vectors $z_{\{0,1,2,3\}}$
into the basis.
Each of the $z_i$ contains four non--overlapping periodic fermions from
the set $\{{\bar\psi}^{1,\dots,5},{\bar\eta}^{1,2,3},{\bar\phi}^{1,\dots,8}\}$.
In the new basis we rename ${\bar\psi}^5\equiv{\bar\eta}^0$.

To illustrate the origin of the spinor--vector duality in the new basis
we will consider first a class of models that are generated by a minimal
set of seven basis vectors, excluding the geometrical coordinate sets $e_i$,
of the basis in eq. (\ref{basis}). The remaining 8 holomorphic
and 32 anti--holomorphic world--sheet fermions are divided into five
non--overlapping groups of eight real fermions.
Such a division in ten dimensions \cite{lewellen1986, fff1,fff2}
is unique and independently of GGSO projection coefficients
always produces in the space--time supersymmetric case either $SO(32)$ or
$E_8\times E_8$, and not any other
gauge groups. Although, naively one may
expect that other gauge groups, like $SO(8)^4$, $SO(16)^2$ or $SO(8)\times
SO(24)$ may arise, the chiral modular properties of the partition function
forbid the other possible extensions in the supersymmetric case.
In terms of the particular $SO(8)$ characters this property follows from
the triality structure of the individual $SO(8)$ character. Namely,
the equivalence of the $8_{_V}$, $8_{_S}$ and $8_{_C}$
$SO(8)$ representations.
This equivalence enables twisted constructions of the $E_8\times E_8$
or $SO(32)$ gauge groups. This phenomena will appear in the models generated
by the new basis introduced below.

The non--holomorphic SUSY breaking vector $b_1$ generates a $Z_2$ projection
which breaks $N=4$ to $N=2$ space--time supersymmetry, and breaks one
of the $SO(8)$ groups to $SO(4)\times SO(4)\equiv SU(2)^4$.

The class of free fermionic models under investigation
is generated by a set $V$ of 7 basis vectors
$$
V=\{v_1,v_2,\dots,v_{7}\},
$$
where
\begin{eqnarray}
v_1=1&=&\{\psi^\mu,\
\chi^{1,\dots,6},y^{1,\dots,6}, \omega^{1,\dots,6}| \nn\\
& & ~~~\bar{y}^{1,\dots,6},\bar{\omega}^{1,\dots,6},
\bar{\eta}^{1,2,3},
\bar{\psi}^{1,\dots,5},\bar{\phi}^{1,\dots,8}\},\nn\\
v_2=S&=&\{\psi^\mu,\chi^{1,\dots,6}\},\nn\\
v_{3}=z_1&=&\{\bar{\phi}^{1,\dots,4}\},\nn\\
v_{4}=z_2&=&\{\bar{\phi}^{5,\dots,8}\},\nn\\
v_{5}=z_3&=&\{\bar{\psi}^{1,\dots,4}\},\nn\\
v_{6}=z_0&=&\{\bar{\eta}^{0,1,2,3}\},\nn\\
v_{7}=b_1&=&\{\chi^{34},\chi^{56},y^{34},y^{56}|\bar{y}^{34},
\bar{y}^{56},\bar{\eta}^0,\bar{\eta}^{1}\},\label{basis2}
\end{eqnarray}
where ${\bar\eta}^0\equiv{\bar\psi}^5$.
The models generated by the basis (\ref{basis2})
preserve $N=2$ space--time supersymmetry, as only one $Z_2$ SUSY breaking projection
is induced by the basis vector $b_1$.
%, has been included in the basis.
The models that preserve only $N=1$ space--time supersymmetry
are easily incorporated by including a second $Z_2$ SUSY breaking projection
given by a second basis vector $b_2$
(see {\it e.g.} ref. \cite{fkr} and references therein).
Here we focus only on the
$N=2$ preserving vacua.
The second function of the second $Z_2$ basis vector $b_2$ is
to break the observable symmetry gauge group from $SO(12)\times SO(4)$ to
$SO(10)\times U(1)^3$. Here the spinor-vector duality is therefore
seen in terms of $SO(12)$, rather than $SO(10)$, representations.

%\subsection{The gauge group}

The gauge groups arising with the new set of
basis vectors eq. (\ref{basis2}).
The sectors contributing to the gauge group are the
$0$--sector and the 10 purely anti--holomorphic sets:
\beqn
G=\{& 0, \nonumber\\
    & z_0,z_1,z_2,z_3, \nonumber\\
    & z_0+z_1,z_0+z_2,z_0+z_3,z_1+z_2,z_1+z_3,z_2+z_3~~\}
\label{gaugesectors}
\eeqn
where the $0$--sector requires two oscillators acting on the vacuum
in the gauge sector; the $z_j$--sectors require one oscillator; and
the $z_i+z_j$ require no oscillators.
We first analyse the $N=4$ gauge group arising prior to the
inclusion of the basis vector $b_1$, which reduces $N=4$ to
$N=2$ space--time supersymmetry. The $b_1$ basis vector does
not give rise to additional enhancement sectors, and therefore
merely reduces the $N=4$ gauge group to a subgroup.

The $0$--sector gauge bosons give rise to the gauge group
\begin{equation}
\left[SO(12)\right]\times SO(8)^4
\label{neq4gg}
\end{equation}
where the $SO(12)$ group factor arises from the 12 right--moving
world-sheet fermions $\{{\bar y},{\bar\omega}\}^{1,\cdots,6}$,
which defines the internal lattice at the free fermionic $SO(12)$
enhanced symmetry point,
and the $SO(8)_{3,0,1,2}$ group factors arise respectively from:
${\bar\psi}^{1,\cdots,4}$, ${\bar\eta}^{0,1,2,3}$,
${\bar\phi}^{1,\cdots,4}$, ${\bar\phi}^{5,\cdots,8}$.
The appearance of the lattice $SO(12)$ gauge group in four dimensions
that can extend the ten dimensional
$SO(32)$ and $E_8\times E_8$ gauge groups, depending on the choices
of GGSO projection coefficients, which may correlate the characters of
the $SO(12)$ lattice characters with those of the four $SO(8)$'s.
The notation used in this division adheres to the conventional
notation used in the free fermionic constructions and the quasi--realistic
heterotic--string models in the free fermionic formulation.

The additional sectors in eq. (\ref{gaugesectors}) can give rise
to space--time vector bosons that enhance the four dimensional
gauge group given in eq. (\ref{neq4gg}). The enhancements depend
on the GGSO phases $\cc{z_i}{z_j}$ with $i\ne j$.
All vacua contain $N=4$ space--times supersymmetry, which fixes
the $\cc{S}{z_i}$ phases. Hence, there may be a priori $2^6$ possibilities
for the four dimensional gauge group, some of which may be repeated.
Identical manifestations of the gauge groups arise from to the
twisted realisation of the group generators, using the triality property of
the $SO(8)$ representations. This is the four dimensional
manifestation of the twisted generation of gauge groups already
noticed in the ten dimensional case.
We list a few of the possibilities in table \ref{tab:gauge}.
\begin{table}
\begin{center}
\begin{tabular}{c c c c c c| r}
$\cc{z_0}{z_1}$ & $\cc{z_0}{z_2}$ & $\cc{z_0}{z_3}$ &
$\cc{z_1}{z_2}$ & $\cc{z_1}{z_3}$ & $\cc{z_2}{z_3}$ &
Gauge group G\\\hline
 +  &  +  &  +  &  +  &  +  &  +  & $SO(44) $\\
$-$ &  +  &  +  &  +  &  +  &  +  & $SO(28) \times E_8 $\\
$-$ & $-$ &  +  &  +  &  +  &  +  & $SO(20) \times SO(24) $\\
 +  &  +  & $-$ & $-$ &  +  &  +  & $SO(12) \times E_8\times E_8 $\\
$-$ &  +  & $-$ & $-$ &  +  &  +  & $SO(12) \times SO(16) \times SO(16)$\\
$-$ & $-$ & $-$ & $-$ & $-$ & $-$ & $SO(12) \times SO(32)$\\
\end{tabular}
\caption{The configuration of the gauge group of the $N=4$ theory.}
\label{tab:gauge}
\end{center}
\end{table}

\subsection{A simple example of the spinor--vector duality}

Including the basis vector $b_1$
reduces $N=4\rightarrow N=2$ space--times supersymmetry.
To illustrate the spinor--vector duality
in the $N=2$ vacua of the new basis,
akin to the spinor--vector duality discussed
in ref. \cite{fkr}, and in section \ref{neq2},
we choose the initial $N=4$ vacuum
with $\left[SO(12)\right]\times SO(16)\times SO(16)$ gauge group.
This case is realised with the GGSO projection coefficient to be:
\beqn
& &\cc{z_0}{z_1}=~~
\cc{z_0}{z_3}=~~
\cc{z_1}{z_2}=\nonumber\\
 &-&\cc{z_0}{z_2}=
-\cc{z_1}{z_3}=
-\cc{z_2}{z_3}=-1
\eeqn
With this choice the additional sectors, beyond the $0$--sector,
giving rise to extra space--time vector bosons are only $z_2$ and $z_3$.
The additional projection induced by the basis vector $b_1$ reduces
the gauge symmetry arising from the $0$--sector to
\beq
\left[SO(8)\times SO(4)\right]_{\cal L}
\times \left[SO(8)_3\times SO(4)\times SO(4)\right]_O\times
\left[SO(8)_1\times SO(8)_2\right]_H
\eeq
The lattice gauge group is reduced to
$\left[SO(8)\times SO(4)\right]_{\cal L}$.
We define as the observable gauge group arising from the $0$--sector
to be
$\left[SO(8)_3\times SO(4)\times SO(4)\right]_O$,
and the 
$\left[SO(8)_1\times SO(8)_2\right]_H$ is
the hidden gauge group. This labelling of observable and hidden gauge groups
will be clarified below. Both observable and hidden sector gauge groups are
enhanced. The hidden gauge group is enhanced to
$\left[SO(16)\right]_H$ due the extra
vector bosons arising from the sector $z_2$.
The extra vector bosons from the sector $z_3$ enhance
the observable $\left[SO(8)_3\times SO(8)_0\right]_O$ 
to $\left[SO(16)\right]_O$ at the $N=4$ level.
While at the $N=2$ level the $b_1$ projection
reduces $\left[SO(16)\right]_O\rightarrow \left[SO(12)\times SO(4)\right]_O
\equiv\left[SO(12)\times SU(2)_0\times SU(2)_1\right]_O$.
Now we are in the position to define the $N=2$ spinor--vector duality
in terms of the $SO(12)$ representations of the observable sector.
Explicitly, the exchange of the vectorial
12 representation of $SO(12)$ with the spinorial 32 representation.
To illustrate the duality we construct two different models in which
these representations are interchanged due to the choices of the
GGSO projection coefficients.

Consider first the choice of the extra phases to be:
%Taking the additional GGSO
%phases to be
\beq
\cc{b_1}{1,z_0}=-\cc{b_1}{S,z_1,z_2,z_3}=-1~.
\label{vectorialphasechoice}
\eeq
This choice defines a model with 2 multiplets in the $(1,2_L+2_R,12,1,2,1)$
and 2 in the
$(8,2_L+2_R,1,2,1,1)$ representations of
$\left[SO(8)\times SO(4)\right]_{\cal L}
\times \left[ SO(12)\times SU(2)_0\times SU(2)_1\right]_{O}\times
\left[SO(16)\right]_H$.
In this case the sectors contributing to the vectorial 12 representation
of $SO(12)$ are the sectors $b_1$ and $b_1 + z_3$, where the sector $b_1$
produces the $(1,2,2)$ representation and the sectors $b_1+z_3$
produces the $(8_{_S},1,1)$ under the decomposition
$\left[SO(12)\right]_O
\rightarrow \left[SO(8)\times SO(4)\right]_O\equiv
\left[SO(8)\times SU(2)\times SU(2)\right]_O$.
All other states are projected out.
Therefore, there are a total of eight multiplets in the vectorial
representation of the observable $SO(12)$ in this model. These
states also transform as doublets of the observable $SU(2)_1$.

The second choice given by
\beq
\cc{b_1}{1,z_0,z_1}=-\cc{b_1}{S,z_2,z_3}=-1
\label{spinorialphasechoice}
\eeq
This choice defines a model with 2 multiplets in the $(1,2_L+2_R,32,1,1,1)$,
and 2 in the $(1,2_L+2_R,1,1,2,16)$, representations of
$\left[SO(8)\times SO(4)\right]_{\cal L}\times
\left[SO(12)\times SU(2)_0\times SU(2)_1\right]_O\times
\left[SO(16)\right]_H$.
In this case the sectors contributing to the spinorial 32 representation
of $\left[SO(12)\right]_O$ are the sectors $b_1+z_0$ and $b_1 + z_3+ z_0$,
where the sector $b_1+z_0$
produces the $(8_{_V},2,1)$ representation and the sectors $b_1+z_3+z_0$
produces the $(8_{_C},1,2)$ under the decomposition
$\left[SO(12)\right]_O\rightarrow
\left[SO(8)\times SO(4)\right]_O\equiv
\left[SO(8)\times SU(2)\times SU(2)\right]_O$.
The sectors contributing to the vectorial 16 representation
of the hidden $SO(16)$ gauge group are the sectors $b_1$ and $b_1 + z_2$,
where the sector $b_1$
produces the $(8_{_V},1)$ representation and the sector $b_1+z_2$
produces the $(1,8_{_C})$ representation
under the decomposition
$\left[SO(16)\right]_H\rightarrow
\left[SO(8)_1\times SO(8)_2\right]_H$.
The hidden 16 representations
transform as doublets of the observable $SU(2)_1$ group.
All other states are projected out.
Therefore, there are a total of eight multiplets in the spinorial 32
representation of the observable $\left[SO(12)\right]_O$ in this model.

We see that in the first example the vectorial 12 representation of the
observable $\left[SO(12)\right]_O$ is constructed as $12 = (8_{_S},1,1)\oplus
(1,2,2)$, while in the second example the spinorials are constructed as
$32 = (8_{_V},2,1)\oplus(8_{_C},1,2)$ under the decomposition
$SO(12)\rightarrow SO(8)\times SU(2)\times SU(2)$.
Due to the triality of the $SO(8)$ representations we may ``rename''
$8_{_S}\leftrightarrow 8_{_V}\leftrightarrow 8_{_C}$.
Renaming thus the $SO(8)$ representations we recover the canonical
decomposition of $SO(n+m)\rightarrow SO(n)\times SO(m)$ as
$V^{n+m}=(V^n,1)\oplus(1,V^m)$,
and
$S^{n+m}=(S^n,S^m)\oplus(C^n,C^m)$,
for the vectorial and spinorial representations of $SO(n+m)$,
respectively\footnote{The vector 4 representation of $SO(4)$ decomposes as
$(2,2)$ under $SU(2)\times SU(2)$.}.
We therefore note that the triality of the $SO(8)$ representations
enables the twisted realisations of the GUT gauge group and representations,
being $SO(12)$ in the $N=2$ models studied here, and $SO(10)$ in $N=1$ models.
This observation offers a novel insight into the realisation of the GUT
symmetries in heterotic string models, and in particular, on
possible relations to other string limits.

The map between the two models,
(\ref{vectorialphasechoice}) and (\ref{spinorialphasechoice}),
is induced by the discrete GGSO
phase change
\ba
\cc{b_1}{z_1}=+1~~\rightarrow~~\cc{b_1}{z_1}=-1
\label{dualitymap}
\ea
Similar to the $x$-map
of refs. \cite{xmap, fkr} the map from
sectors that produce vectorial representations of the observable $SO(12)$
group to sectors that
produce spinorial representations in the models utilising the
basis of eq. (\ref{basis2}) is obtained by adding
the basis vector $z_0$.
Appropriate choice of the discrete GGSO phases can
project the vectorial states and maintain the spinorial states
and visa versa.
The discrete phase change from
(\ref{vectorialphasechoice}) to (\ref{spinorialphasechoice})
indeed induces the spinor--vector duality map in the $N=2$ model.
The role of the basis vectors $z_2$ and $z_3$ in the models of
(\ref{vectorialphasechoice}) and (\ref{spinorialphasechoice})
is to generate
the twisted realisation of the  gauge symmetry enhancement
of the $SO(8)$ gauge groups arising from the null sector.
The space of $N=2$ free fermionic
heterotic string models, which is generated by the basis (\ref{basis})
can now be spanned by adding the $e_i$ vectors to (\ref{basis2}).
Similarly, the space of $N=1$ vacua classified in \cite{fknr,fkr}
can be generated by supplementing (\ref{basis2}) with the second
$Z_2$ breaking vector $b_2$.

\section{Conclusions}\label{conclude}

In this paper we demonstrated that the spinor--vector duality
observed in ref. \cite{fkr} in $Z_2\times Z_2$ free fermionic $N=1$
space--time supersymmetric vacua exists also in
$Z_2$ free fermionic $N=2$ vacua, {\it i.e.} prior
to the inclusion of a second supersymmetry breaking $Z_2$ twist.
In the case of $N=2$ vacua the duality map is between the total number
of (32,1,1) and (32$^\prime$,1,1) spinorial representations of the
observable $SO(12)\times SU(2)\times SU(2)$ versus the total number
of (12,2,1) and (12,1,2) vectorial representations. The $N=2$ vacua
contain a single twisted sector, which facilitates the algebraic proof
given in section \ref{proof}.

We further demonstrated the duality by introducing a novel basis, eq.
(\ref{basis2}) to generate the free fermionic models. The earlier basis,
used in section \ref{neq2}
%The notation used in the basis of eq. (\ref{basis})
follows the usual division used in the literature of
the quasi--realistic free fermionic models. This division reflects the two
key characteristics of these models, being: a) their relation to $Z_2\times
Z_2$ orbifold compactifications; b) the $SO(10)$ GUT symmetry generated
by the complex world--sheet fermions ${\bar\psi}^{1\cdots5}$.
In the new basis given in eq. (\ref{basis2}) the underlying $SO(10)$ symmetry
is no longer manifest. An $SO(10)$ GUT symmetry can arise with the
new basis for appropriate choices of the GGSO phases due to enhancements
from additional sectors. This is an important distinction for several
reasons:

1. The rank 16 gauge group is now even more symmetric. The gauge
degrees of freedom are grouped into four groups of four, each generating
an $SO(8)$ modular block.
The enhancement of one $SO(8)\times SO(2)$ (or $SO(8)\times U(1)$)
to $SO(10)$ is obtained
for a particular choice of the GGSO phases. But any one of three $SO(8)$s
can be enhanced. In fact, all three $SO(8)$s can be enhanced simultaneously
yielding a model with $SO(10)^3$ gauge symmetry.
The $SO(10)$ symmetry is generated by grouping states
from the trivial and non--trivial sectors.
The character of the $SO(10)$ representations, {\it i.e.} whether it
spinorial or vectorial and its chirality, resides in the $U(1)$ charges.
The new manner in which
the $SO(10)$ symmetry is obtained may shed new light on the origin
of the GUT symmetries in heterotic string theory and its relation to other
limits.

2. The new division of the world--sheet fermions of the rank 16
gauge degrees of freedom is well known in the classification of the
ten dimensional heterotic string of ref. \cite{lewellen1986, fff1, fff2}.
The only supersymmetric vacua in 10 dimensions are the $SO(32)$ and
$E_8\times E_8$ vacua. However, there are different ways to generate
this symmetry in terms of the basis generators $\{z_0,z_1,z_2,z_3\}$
of (\ref{basis2}), all of which produce equivalent symmetries. This
is a reflection of the triality property of the $8_{_V}$, $8_{_S}$ and
$8_{_C}$ $SO(8)$ representations.
In four dimensions other enhancements are possible due to the symmetries
arising from the compactified six dimensional lattice at the
enhanced symmetry point, which is reflected in table \ref{tab:gauge}.

3. The $Z_2$ supersymmetry breaking vector $b_1$ in (\ref{basis2})
breaks one and only one of the four untwisted $SO(8)$s to $SO(4)\times SO(4)$.
One of the remaining $SO(8)$s may be combined with one of these $SO(4)$s
to form an $SO(12)$ symmetry group. The triality characteristic of the
enhanced $SO(8)$ representation is lost, as is now reflected in the
spinor--vector duality. Thus the spinor--vector duality has it roots in
the modular properties of the original $SO(8)$ modular blocks.

4. To date the heterotic $E_8\times E_8$ and occasionally the heterotic
$SO(32)$ has held a special position in terms of attempts to relate string
vacua to experimental particle data. The results of this paper, however,
suggest that this supreme position should be examined, and that the
more fundamental role may be played by the $SO(8)$ characters. This view
opens up many interesting questions for investigation.

\section{Acknowledgements}

AEF would like to thanks the Oxford theory department for hospitality
and is supported in part by STFC under contract PP/D000416/1.
CK is supported in part by the EU under the contracts
MRTN-CT-2004-005104, MRTN-CT-2004-512194,
MRTN-CT-2004-503369, MEXT-CT-2003-509661, CNRS PICS 2530, 3059 and
3747, ANR (CNRS-USAR) contract  05-BLAN-0079-01.
JR work is supported in part by the EU under contracts
MRTN--CT--2006--035863--1 and  MRTN--CT--2004--503369.

%=========================================================================
%======================== REFERENCES =====================================
%=========================================================================

%\vfill\eject

\bigskip
\medskip

\bibliographystyle{unsrt}

\end{document}